\documentclass[preprint,12pt]{elsarticle}

\usepackage{amsmath, amsfonts, amssymb}
\usepackage{bm}
\usepackage{graphicx}
\usepackage{hyperref}
\usepackage{soul}
\usepackage{color}
\usepackage{comment}

\enlargethispage{\baselineskip}

\hypersetup{colorlinks   = true,
            urlcolor     = blue,
            citecolor    = blue,
            linkcolor    = blue,
            menucolor    = blue,
            anchorcolor  = blue,
            filecolor    = blue
}
\everymath{\displaystyle}

\journal{Physics of the Dark Universe}

\begin{document}
\begin{frontmatter}

\title{Do neutrinos bend? Consequences of an ultralight gauge field as dark matter}

\author[a, b]{Luca Visinelli\corref{cor1}}
\ead{luca.visinelli@sjtu.edu.cn}
\author[a,c]{Tsutomu T.\ Yanagida}
\ead{tsutomu.tyanagida@sjtu.edu.cn}
\author[a,b]{Michael Zantedeschi}
\ead{zantedeschim@sjtu.edu.cn}
\cortext[cor1]{Corresponding author}

\address[a]{Tsung-Dao Lee Institute (TDLI), No.\ 1 Lisuo Road, 201210 Shanghai, China}
\address[b]{School of Physics and Astronomy, Shanghai Jiao Tong University, \\ Dongchuan Road 800, 200240 Shanghai, China}
\address[c]{Kavli IPMU (WPI), The University of Tokyo, Kashiwa, Chiba 277-8583, Japan}

\date{\today}

\begin{abstract}
An ultralight gauge boson could address the missing cosmic dark matter, with its transverse modes contributing to a relevant component of the galactic halo today. We show that, in the presence of a coupling between the gauge boson and neutrinos, these transverse modes affect the propagation of neutrinos in the galactic core. Neutrinos emitted from galactic or extra-galactic supernovae could be delayed by $\delta t = \left(10^{-8}\textrm{--}10^1\right)$\,s for the gauge boson masses $m_{A'} = \left(10^{-23}\textrm{--}10^{-19}\right)$\,eV and the coupling with the neutrino $g= 10^{-27}\textrm{--}10^{-20}$. While we do not focus on a specific formation mechanism for the gauge boson as the dark matter in the early Universe, we comment on some possible realizations. We discuss model-dependent current bounds on the gauge coupling from fifth-force experiments, as well as future explorations involving supernovae neutrinos. We consider the concrete case of the DUNE facility, where the coupling can be tested down to $g \simeq 10^{-27}$ for neutrinos coming from a supernova event at a distance $d = 10$\,kpc from Earth.
\end{abstract}

\flushbottom
\end{frontmatter}

\section{Introduction}

Dark matter (DM) could be made up of a light boson field~\cite{Preskill:1982cy,Abbott:1982af,Dine:1982ah}: examples include the Nambu-Goldstone boson~\cite{Nambu:1960tm,Goldstone:1961eq} such as the QCD axion~\cite{Weinberg:1977ma, Wilczek:1977pj}, the majoron~\cite{Chikashige:1980ht, Gelmini:1980re, Liang:2024vnd}, or a massive hidden photon~\cite{Okun:1982xi,Georgi:1983sy}. Motivations for an ultralight gauge boson can also be found in Refs.~\cite{Fayet:1980ad, Fayet:1989mq}. Here, we focus on the case in which DM is in the form of hidden photons, an attractive cold DM candidate for which various production mechanisms have been explored in the literature~\cite{Redondo:2008ec, Nelson:2011sf, Arias:2012az, Fradette:2014sza, An:2014twa, Graham:2015rva, Agrawal:2018vin, Dror:2018pdh, Co:2018lka, Bastero-Gil:2018uel, Long:2019lwl, Fabbrichesi:2020wbt, Caputo:2021eaa}. 

For example, there are motivations to speculate on the case for a $U(1)_{B\textrm{--}L}$ gauge symmetry as being natural and well motivated. Gauge anomalies are canceled out by introducing three right-handed neutrinos $N_i$, where $i\in\{1,2,3\}$ denotes the neutrino generation, which in turn acquire large Majorana masses from the spontaneous symmetry breaking. The presence of heavy Majorana neutrinos is key for inducing the small masses of the active neutrinos through the seesaw mechanism~\cite{Minkowski:1977sc, Yanagida:1979as, Yanagida:1979gs, GellMann:1980vs, Wilczek:1979hh, Mohapatra:1979ia} and generating the baryon asymmetry in the Universe via the leptogenesis mechanism~\cite{Fukugita:1986hr, Buchmuller:2005eh}.

The spontaneous breaking of the $U(1)_{B\textrm{--}L}$ gauge symmetry is accompanied by the origin of a massive $B\textrm{--}L$ gauge boson which could itself be the DM. Various mechanisms for producing the $B\textrm{--}L$ gauge boson as the dominant DM component have been explored involving gravitational production during reheating~\cite{Graham:2015rva}, misalignment during inflation~\cite{Graham:2015rva, Nakayama:2019rhg, Nakai:2022dni}, a dark thermal bath~\cite{Choi:2020dec}, resonant conversion of an axion field or a dark Higgs~\cite{Agrawal:2018vin, Co:2018lka, Bastero-Gil:2018uel, Dror:2018pdh}, and the decay of the cosmic string network associated with the breaking of the $U(1)_{B\textrm{--}L}$ gauge symmetry~\cite{Long:2019lwl, Kitajima:2022lre}; this latter mechanism is associated with additional features such as a characteristic strain of gravitational waves (GWs)~\cite{Vilenkin:1986ku, Vilenkin:2000jqa}, in addition to the release of the gauge bosons~\cite{Hindmarsh:2008dw, Hiramatsu:2020zlp, Kitajima:2022lre}. Another potential target for a multi-messenger search lies in the low-energy neutrinos released by the decay of the gauge boson, owing to the potentially suppressed decay rate into three photons~\cite{Lin:2022xbu, Lin:2022mqe, Choi:2020kch, Okada:2020evk, Sheng:2023iup, Hooper:2023fqn}.

Besides the direct search for new gauge bosons and GWs, the existence of a massive gauge boson in the DM halo could be inferred indirectly from the properties of neutrinos propagating over galactic distances. Astrophysical neutrinos represent one of the main targets for future theoretical explorations as they could carry information on new physics~\cite{Ackermann:2022rqc}. For this, the search is undergoing or planned at various facilities worldwide. The IceCube collaboration~\cite{IceCube:2014stg, IceCube:2020acn, IceCube:2021rpz} have reported the detection of an approximately isotropic flux of astrophysical neutrinos, covering energies between the TeV and several PeV. Near-future detectors will increase the sensitivity in energy and angular resolution, including the next generation of IceCube~\cite{IceCube-Gen2:2020qha, IceCube-Gen2:2021rkf}, the Cubic Kilometre Neutrino Telescope or KM3NeT 2.0~\cite{KM3Net:2016zxf, KM3NeT:2018wnd}, The tRopIcal DEep-sea Neutrino Telescope (TRIDENT)~\cite{Ye:2022vbk}, and the Pacific Ocean Neutrino Experiment (P-ONE)~\cite{P-ONE:2020ljt, Resconi:2021ezb}. Supernova neutrinos will be studied at facilities such as Hyper-Kamiokande~\cite{Hyper-Kamiokande:2021frf}, the Jiangmen Underground Neutrino Observatory (JUNO)~\cite{JUNO:2015zny}, and the Deep Underground Neutrino Experiment (DUNE)~\cite{DUNE:2015lol}. For recent reviews of the expected features in neutrino signals see Refs.~\cite{Mirizzi:2015eza, Horiuchi:2018ofe}.

The propagation of neutrinos over cosmological distances can be affected by various effects such as strong gravitational lensing~\cite{Taak:2022xdp} and the scattering of TeV neutrinos off dark matter~\cite{Koren:2019wwi, Bustamante:2020mep}. Here, we point out that another source for the deflection of a neutrino beam resides in its interaction with the transverse component of the field associated with a vector DM model. While we do not specifically consider a particle model, we comment on the $B\textrm{--}L$ gauge boson as a concrete example. We list the signatures that could carry a potential fingerprint for the existence of this model such as the deflection of the neutrino beam, the time delay with respect to a companion electromagnetic or GW signature, and the directionality of the signal. Our findings are independent of the actual mechanism leading to the production of the light vector DM field, due to the late-time galactic dynamics ensuring the presence of transverse modes in the halo~\cite{Amaral:2024tjg}.

We anticipate that the model proposed might result in a time delay of a neutrino beam from a nearby supernova in the range $\delta t = \left(10^{-8}\textrm{--}10^1\right)$\,s for the gauge boson masses $m_{A'} = \left(10^{-23}\textrm{--}10^{-19}\right)$\,eV and the coupling with the neutrino $g= 10^{-27}\textrm{--}10^{-20}$. While our findings hold for a generic gauge coupling $g$, part of the parameter space is excluded by fifth force experiments when a $B{\rm -}L$ gauge field is explicitly considered. For this reason, a more natural model for the effect we discuss at macroscopic scales is provided when considering a $L_{\mu}\textrm{--}L_{\tau}$ symmetry. Other gauge symmetries, such as $L_{e}\textrm{--}L_{\tau}$ and $L_{e}\textrm{--}L_{\mu}$, can be probed in neutrino oscillation experiments~\cite{Grifols:2003gy, Joshipura:2003jh}, space missions~\cite{Touboul:2017grn, MICROSCOPE:2022doy}, torsion balance experiments~\cite{Adelberger:2009zz,Wagner:2012ui}, and astrometric measurements~\cite{Tsai:2023zza}. In this mass range, the vector gauge boson can be a candidate for ultralight DM, see e.g.~\cite{Agrawal:2018vin}, along with other ``fuzzy'' DM candidates~\cite{Hu:2000ke, Hui:2016ltb}.

We organize the paper as follows. In Sec.~\ref{sec:methods} we introduce the model considered for the gauge boson and the production of ultra-light vector bosons as the DM. Section~\ref{sec:neutrino} discusses the propagation of neutrinos in the galaxy. The methods used for the analysis and the discussion of the results are presented in Sec.~\ref{sec:results}. Conclusions are drawn in Sec.~\ref{sec:conclusions}. In this article, we set $\hbar=c=\epsilon_0=1$.

\section{Light gauge bosons as dark matter}
\label{sec:methods}

To elucidate our findings, we consider the effective Lagrangian
\begin{equation}
    \label{eq:Lagrangian}
    \mathcal{L} = - \frac{1}{4}(F')^{\mu\nu}F'_{\mu\nu} - \frac{1}{2}m_{A'}^2(A')^\mu A'_\mu + g\bar\nu_L\gamma^\mu\nu_L A'_\mu\,,   
\end{equation}
which might arise from an Abelian Higgs model for the coupling of a complex scalar field with the associated gauge boson $A'_\mu$ and field strength $F'_{\mu\nu} = \partial_\mu A'_\nu-\partial_\nu A'_\mu$. Here, $m_{A'}$ is the mass of the gauge boson and $g$ is the gauge coupling with the left-handed neutrino $\nu_L$. Such a coupling can arise, for example, in a $B\textrm{--}L$ or $L_{\mu}\textrm{--}L_{\tau}$ symmetry-breaking theory. In the latter case, care should be used in writing the coupling in Eq.~\eqref{eq:Lagrangian}, which becomes generation dependent. In these UV realizations, the vacuum expectation value $v'$ of a scalar field $\phi$ spontaneously breaks the symmetry and generates the gauge boson mass $m_{A'}\sim g\, v'$. In the $B\textrm{--}L$ symmetric case, the field $\phi$ generates a mass term $m_N\sim y_N\,v'$ for the right handed neutrino $N$ provided it has the proper lepton charge, where $y_N$ is the Yukawa coupling. This, in turn, sets up the seesaw mechanism, leading to neutrino mass $m_{\nu}\sim m_D^2/m_N$. In the $L_{\mu}\textrm{--}L_{\tau}$ symmetric case, further care is required when considering these operators; we refer the interested reader to Ref.~\cite{Granelli:2023egb} for details.

The Lagrangian of the gauge field in Eq.~\eqref{eq:Lagrangian} leads to the equation of motion in unitary gauge,
\begin{equation}
    \label{eq:maxwell}
    \partial^\mu F'_{\mu\nu} - m_{A'}^2\,A'_\nu = 0\,,
\end{equation}
which implies the constraint $\partial^\mu A'_\mu = 0$. In models where the relic density of gauge bosons constitutes the DM, the vector field is expected to be temporally dependent and oscillating with the frequency $m_{A'}$. The ultralight boson field organizes into a solitonic condensate with a spatially varying radial profile, being coherent over a length scale $R_{\rm core}$ of galactic size. This impacts the structure of the vector field inside the soliton, as the constraint following Eq.~\eqref{eq:maxwell} implies the relation
\begin{equation}
    \label{eq:a0}
    A'_0 \sim \frac{2\pi}{m_{A'}\,R_{\rm core}}|{\bf A'}| \ll |{\bf A'}|\,,
\end{equation}
so that the zeroth component of the vector field can be neglected. The last step above derives from $m_{A'}\,R_{\rm core} \sim v^{-1}$, where $v\sim 10^{-3}$ is the velocity dispersion in the soliton. The spatial component of the vector field is related to its present density $\rho_{A'}$ as $|{\bf A'}| = \sqrt{2\rho_{A'}}/m_{A'}$ so that the electric field is
\begin{equation}
    \label{eq:Efield}
    E' = |{\bf E}'| = \sqrt{2\rho_{A'}}\,.
\end{equation}
This corresponds to $E' \sim 10^{-1}{\rm\,eV^2}$ when the DM is in the form of the ultralight gauge boson. At the same time, the magnetic field ${\bf B}' = \nabla\times {\bf A'}$ is suppressed according to Eq.~\eqref{eq:a0} and can be neglected onwards.

The allowed space is constrained by requiring that the cold population of gauge bosons is not overly produced. If one of the three active neutrinos is massless, the ultralight gauge boson would mainly decay into left-handed neutrinos. If the gauge boson mass is larger than one of the neutrinos it couples to, $m_{A'}\gtrsim m_{\nu}$, its lifetime is given by~\cite{Palomares-Ruiz:2007egs}
\begin{equation}
    \label{eq:tau}
    \tau_{A'} = \frac{8\pi}{g^2m_{A'}}\,,
\end{equation}
so that requiring the boson to be stable over a cosmological timescale $\tau_{A'} > t_0$ with $t_0 = 13.7\,$Gyr leads to the excluded region
\begin{equation}
    g \lesssim 2 \times 10^{-16}\,\left(\frac{\rm eV}{m_{A'}}\right)^{1/2}\,,
\end{equation}
which is easily accommodated for an ultralight boson. Note, that for this bound to be applicable in our relevant range of parameters, an ultralight neutrino is required, otherwise the gauge boson is stable over cosmological timescales.

Various mechanisms that lead to a relic density of gauge bosons with a light mass $m_{A'}$ have been proposed in the literature. One such example includes the parametric resonant production from axion decay~\cite{Agrawal:2018vin} or from the decay of the inflaton field~\cite{Bastero-Gil:2018uel}. More generally, as shown below, the model proposed leads to a symmetry breaking scale $v'\sim m_{A'}/ g\sim \mathcal{O}({\rm eV}\textrm{--}{\rm MeV})$. If the symmetry considered is $B\textrm{--}L$, strong constraints follow on the gauge coupling from equivalence principle (EP) experiments~\cite{Adelberger:2009zz,Wagner:2012ui}, which impose $g \lesssim 10^{-24}$. These constraints exclude part of the parameter space where the bending of neutrinos leads to detectable effects. One possible way to avoid this is to consider theories based on different symmetries.
One such model has been recently built in the context of a minimal gauge symmetry by introducing a gauged Abelian model preserving $L_{\mu}\textrm{--}L_{\tau}$ to address thermal leptogenesis~\cite{Granelli:2023egb}. Similar models could also be potentially built by considering a $B\textrm{--}L$ gauged symmetry for the 2nd or 3rd generations, which would partially relax EP constraints. 
Regardless of the symmetry considered, such a low symmetry breaking scale requires additional care, if the relic density of gauge boson is to be the observed DM. A possible model, as already mentioned, is offered in~\cite{Agrawal:2018vin}. In this case, the gauge vector bosons are produced via decay of an axion field and it is compatible with the ``fuzzy'' DM mass range relevant for the present mechanism.

Note, that ultralight bosons as DM can be probed efficiently from the observation of spinning supermassive BHs, which implies the absence of such a boson with the Compton wavelength of comparable size as the Schwarzschild radius~\cite{Arvanitaki:2010sy}. For the ultralight gauge boson field, the estimate of the spin parameter of the BH M87$^\star$ by the Event Horizon Telescope collaboration excludes the mass range around $\sim 10^{-21}$\,eV~\cite{Davoudiasl:2019nlo, Chen:2019fsq}. More massive BHs are theoretically possible and could lead to excluding the existence of even lighter particles~\cite{Carr:2020erq}.

Given the production of a coherent state of light vector bosons, gravitational interactions generally predict that the transverse and longitudinal modes admixture to about similar amounts~\cite{Amaral:2024tjg}. After their production, the quanta have kinetic energy of the order of $m_{A'}$ and therefore rapidly turn non-relativistic due to cosmological redshift. The gauge boson distribution is expected to be spatially dependent, with $\mathcal{O}(1)$ overdensities within which the gauge boson forms coherent patches with a correlation length of the order of $\lambda_{\rm in}\sim m_{A'}^{-1}$. The time and space variations of the field strength then seeds an electric and a magnetic fields that is coherent over the scale $R_{\rm core}$ at the present epoch. Within a given patch, the gauge field is coherent and effectively corresponds to a background electric field of amplitude $E'$ as in Eq.~\eqref{eq:Efield}.

\section{Neutrino propagation}
\label{sec:neutrino}

The effective electric field in Eq.~\eqref{eq:Efield} is homogeneous and approximately constant on length scales of the order of $R_{\rm core}$. Within this region, all neutrinos experience an electric field strength inside the soliton, with a spatial distribution that depends on the radial profile of the soliton. In more detail, a neutrino of energy $\epsilon_\nu$ experiences the acceleration $a_\nu \simeq g\,E_{\perp}/\epsilon_{\nu}$, where $E_{\perp}$ denotes the component of the ``effective'' electric field that is perpendicular to the direction of motion. This leads to a deflection over the soliton by an angle
\begin{equation}
    \label{eq:Theta}
    \Theta \simeq \frac{g\,E_{\perp}}{2\epsilon_\nu}\,\frac{L^2}{d}\,,
\end{equation}
where $d$ is the distance from the source and $L < d$ is the length of the path spent by the neutrino in the soliton. Correspondingly, the neutrino beam experiences a deviation by a length $y = d\,\Theta$. Note, that the coupling $g$ above differs from the gauge coupling $g$ appearing in the Lagrangian of Eq.~\eqref{eq:Lagrangian}, due to the modifications by the unitary matrix rotating neutrinos to Majorana mass eigenstates. Nevertheless, the two couplings are of the same order of magnitude except for specific values of the neutrino mixing angles. We neglect such corrections from here onwards. Moreover, potentially dangerous lepton number violating processes of the form $\nu \rightarrow \bar{\nu}$ are negligible as their probability is helicity suppressed approximately as $\left(m_{\nu}/\epsilon_{\nu}\right)^2$~\cite{Schechter:1980gk,Li:1981um,Bernabeu:1982vi,Langacker:1998pv,deGouvea:2002gf}. We also neglect effects from neutrino electromagnetic interactions~\cite{Lambiase:2004qk, MosqueraCuesta:2008skn} or a curved background~\cite{Stodolsky:1978ks, Visinelli:2014xsa}.

We obtain the length travelled inside the soliton as
\begin{equation}
    \label{eq:length}
    L = 2\,(R_{\rm core}^2 - r_\odot^2\sin^2\psi)^{1/2}\,,    
\end{equation}
where $\psi$ is the angle between the trajectory of the neutrino and the axis containing the Galactic center and the Solar system, and $r_\odot \simeq 8.5$\,kpc. The angle $\psi$ is related to the right ascension and declination in Galactic coordinates $(\ell, b)$ as $\cos\psi = \cos\ell\cos b$. Future events constraining this model demand that the neutrino signal crosses the soliton core. Therefore, the direction of the event should obey
\begin{equation}
    \label{eq:psiangleconstr}
    |\sin\psi| < R_{\rm core}/r_\odot\lesssim 1/10\,.    
\end{equation}
Here, we work under such an assumption. Note, that since the thickness of the Milky Way is of the order of $\mathcal{O}(1\,\textrm{kpc})\ll r_{\odot}$, the angle $\ell$ needs to obey the condition in Eq.~\eqref{eq:psiangleconstr} for a galactic supernova. Moreover, the distance of the supernova must be larger than $r_{\odot}$ for the neutrino to cross the central galactic region before reaching Earth. On the other hand, for a supernova emitting from an isotropic extra-galactic population, the angle $b$ would also be constrained as in Eq.~\eqref{eq:psiangleconstr} leading to a probability $\propto |\sin\psi|^2$. For example, the SN1987A event itself cannot be used to directly place a bound over this model, given that the SN event of Galactic coordinates $\ell = 279.7^\circ$, $b = -31.9^\circ$ does not align with the soliton core.

We model the soliton core as follows. Simulations predict that the central core structure of a soliton formed within a halo of mass $M_h$ is well described by the expression~\cite{Schive:2014dra, Schive:2014hza}
\begin{equation}
    \label{eq:rhosoliton}
    \rho_{A'}(r) = \frac{\rho_c}{[1 + \gamma (r/r_c)^2]^8}\,,
\end{equation}
with the comoving core density and the cutoff radius $r_c$ given by~\cite{Schive:2014hza, Taruya:2022zmt}
\begin{equation}
    \label{eq:correlation}
    \begin{split}
    \rho_c = 0.019\,M_\odot{\rm\,pc}^{-3}\,\left(\frac{10^{-22}\,\rm eV}{m_{A'}}\right)^{2}\!\left(\frac{\rm kpc}{r_c}\right)^4\,,\\
    r_c = 1.6{\rm\,kpc}\,\left(\frac{10^{-22}{\rm\,eV}}{m_{A'}}\right)\,\left(\frac{M_h}{10^9\,M_\odot}\right)^{-1/3}\,.
    \end{split}
\end{equation}
The constant $\gamma \approx 0.0905$ is fixed by demanding that the density drops to one-half of its central density at the radius $r_c$, as $\rho_{A'}(r_c) = \rho_c /2$. Although $r_c$ can be of astronomical scale for the case of an ultralight particle, this parameter is generally much smaller than the distance $d$ at which the neutrino is sourced through some astrophysical process. Here, we fix $r_c = 160{\rm\,pc}\,\left(10^{-22}{\rm\,eV}/m_{A'}\right)$ given the size of the Milky Way halo.

We define the size of the core $R_{\rm core}$ by setting $\rho_{A'}(R_{\rm core}) = \rho_{\rm NFW}(R_{\rm core})$, where $\rho_{\rm NFW}(r)$ is the Navarro-Frenk-White (NFW) profile~\cite{Navarro:1995iw} with $r_s = 16\,$kpc. This leads to $R_{\rm core} \approx (10\textrm{--}1000)$\,pc, depending on the gauge boson mass. In the numerical results below, we have accounted for the fact that the distribution of the gauge boson in the soliton is not uniform, being described by the profile in Eq.~\eqref{eq:rhosoliton} for $r < R_{\rm core}$.

Deflection also implies a time delay compared to the straight propagation given by
\begin{equation}
    \label{eq:timedelay}
    \delta t \simeq \frac{y^2}{2L} \simeq \frac{g^2}{4\epsilon_\nu^2}\,L^3\,\bar\rho_{A'}\,,
\end{equation}
where $\bar\rho_{A'}$ is the average DM density within $R_{\rm core}$. Contrary to the deflection angle $\Theta$, the time delay depends on the size of the traversed solitonic core $L$ while being independent on the distance from the neutrino source $d$. The parallel component of the electric field also plays a role by accounting for the spreading of the arrival time, with a standard deviation that is comparable to the result in Eq.~\eqref{eq:timedelay}.

As an example, consider a neutrino beam of average energy $\bar\epsilon_\nu \approx 15$\,MeV, sourced by a supernova at a distance $d \sim \mathcal{O}(100{\rm\,kpc})$ which is of the same order as SN1987A. If the neutrinos traverse a region containing a coherent patch of gauge bosons of size $L \sim {\rm kpc}$, both the deflection angle and the time delay depend on the gauge boson density $\rho_{A'}$, which is here assumed to be distributed according to Eq.~\eqref{eq:rhosoliton}. The deflection from the source and the time delay would approximately amount to
\begin{equation}
    \begin{split}
    \Theta &\approx 3\times 10^{-5}\,\left(\frac{g}{10^{-22}}\right)\left(\frac{L^2}{d{\rm\,kpc}}\right)\left(\frac{m_{A'}}{10^{-22}{\rm\,eV}}\right)^{0.83},\\
    \label{eq:timedelaynum}
    \delta t &\approx 30{\rm\,s}\,\left(\frac{g}{10^{-22}}\right)^2\,\left(\frac{L}{\rm kpc}\right)^4\,\left(\frac{m_{A'}}{10^{-22}{\rm\,eV}}\right)^{1.65},
    \end{split}
\end{equation}
where the numerical exponents in $m_{A'}$ account for the fit of the average density $\bar\rho_{A'}$ with the gauge boson mass. In the case where the neutrino burst originates at the Galactic coordinates $(\ell, b)$, the beam would experience the length path $L$ as in Eq.~\eqref{eq:length} with $\cos\psi = \cos\ell\cos b$. This, jointly with the measured delay between photons and neutrinos $\Delta t_{\gamma \nu}$, would then constrain the gauge coupling $g$.

For reference, we focus on the upper bound obtained for an ideal signal crossing the Galactic center by setting $\delta t \lesssim \Delta t_{\gamma \nu}$, where $\Delta t_{\gamma \nu} \sim 10{\rm\,s}$ is the bound from the neutrino burst released during the SN1987A event~\cite{Arnett:1987iz, Kolb:1987dda, Raffelt:1996wa}. This results in an upper bound on the gauge boson coupling $g$ for a given mass $m_{A'}$. For example, this gives the bound $g \lesssim 10^{-21}$ for $m_{A'} = 10^{-22}\,$eV.

The effect described competes with the delay due to the finite mass of the neutrino
\begin{equation}
    \label{eq:deltatnu}
    \delta t_\nu = \frac{m_\nu^2}{2\epsilon_\nu^2}\,d\,,
\end{equation}
whose dependence on the distance from the source scales as $d$, as opposed to the effect of the motion of the neutrinos in the DM soliton in Eq.~\eqref{eq:timedelay} which does not depend on $d$. As such, this difference allows for a discrimination between the two competing phenomena. In addition to this, neutrinos coming from different regions in the sky would experience a variable effective length while travelling inside the coherent regions and its associated densities, so that the time delay effect depends on the direction of propagation of the neutrinos with respect to the Galactic center.

For the case of an individual supernova event, demanding that the effect is visible in future experiments leads to the bound $\delta t > \delta t_\nu$, or
\begin{equation}
    \label{eq:distancenu}
    d \lesssim g^2\,L^3\,\frac{\rho_{A'}}{2 m_\nu^2}\,,
\end{equation}
where here we use the lowest bound from the sum of the three active neutrinos $m_\nu \gtrsim 0.06$\,eV resulting from the analysis of neutrino oscillations~\cite{Forero:2012faj, Capozzi:2013psa, ParticleDataGroup:2022pth}. Since we expect $d \gtrsim\,$kpc for the effect to take place over a considerable fraction of the cored halo, a necessary condition on the gauge boson coupling so that a visible effect is produced is $g \gtrsim 10^{-26}$ for $m_{A'}=10^{-22}$\,eV, as we obtain below. A different competing effect is the influence of the mean galactic field, which also affects neutrino species differently. Assuming a relativistic neutrino, the acceleration due to the soliton's presence is approximately $a_g \sim GM_h/R_{\rm core}^2$, and this must be compared to the acceleration from the gauge boson field, $a \sim g E'|/\epsilon_\nu$. Gravity becomes significant when $g \lesssim G \sqrt{2\rho_{A'}} R_{\rm core} \epsilon_\nu \sim 10^{-26}$, which impacts the lower portion of the results we present in the following section. These limitations can be circumvented by detecting neutrino beams coming from different directions of the sky or from different distances given the different dependence over $d$ and $L$ in the two time delay formulas.

One crucial detail resides in the screening of the gauge boson field due to the streaming of cosmic neutrinos. If the gauge boson is the DM, neutrinos organize inside a coherent patch by configuring a screening electric field $E_{\rm screen} = 2R_{\rm core}\,g\,n_\nu$, where the number density of neutrinos is estimated in terms of the baryon-to-photon ratio at recombination $\eta \approx 6.2\times 10^{-10}$~\cite{Planck:2018vyg} and the baryon density $n_b$ as $n_\nu \sim n_b/\eta \approx 410{\rm\,cm}^{-3}$. Screening is negligible for $E_{\rm screen} <E'$, or
\begin{equation}
    \label{eq:gblscreening}
    g < \frac{\sqrt{2\bar\rho_{A'}}}{2\,R_{\rm core}\,n_\nu}\,.
\end{equation}
This relation effectively constraints the viable parameter space region, as discussed below. Note, that the violation of the bound in Eq.~\eqref{eq:gblscreening} would lead to the redistribution in phase space of the cosmic neutrino background in the region towards the Galactic center. In Eq.~\eqref{eq:gblscreening} we have employed the expression for the DM density in the core as given in Eq.~\eqref{eq:rhosoliton} at $r=R_{\rm core}$, which also depends on the gauge boson mass. For this, the bound is not expressed in terms of the combination $g/m_{A'}$ as it would be na\"ively expected. Note, that since $R_{\rm core}$ is typically smaller than $r_{\odot}$, the above screening would bear no observational consequences on the cosmic neutrino background surrounding us.

Given the background electric field, a natural question is whether spontaneous pair production of neutrinos can screen it. The Schwinger pair production occurs, approximately, with a rate per unit volume
\begin{equation}
    \Gamma \simeq (g\,E')^2\exp\left(-\frac{m_\nu^2}{g\,E'}\right)\,.    
\end{equation}
The small gauge coupling $g$ generally leads to an exponential suppression of the nucleation process if all neutrino masses are similar and of the order of $m_{\nu}\sim 10^{-2}$\,eV. However, in the limit of one vanishing neutrino mass, the electric field would effectively be screened within the scale $R_{\rm core}$ over a cosmological timescale. This would lead to a broad distribution in the neutrino energies, with a maximum value $\epsilon_{\nu}^{\rm max}\sim \sqrt{g\, E' }\sim 10^{-12}\,(g/ 10^{-22})^{1/2}$\,eV. Unfortunately, these low energies are well below the current threshold expected in cosmic neutrino experiments such as the Princeton Tritium Observatory for Light, Early-Universe, Massive-Neutrino Yield (PTOLEMY)~\cite{PTOLEMY:2018jst, PTOLEMY:2019hkd}.

Before proceeding to the next section, one could ask if the magnetic field, while significantly smaller in magnitude due to the suppression term in Eq.~\eqref{eq:a0}, could lead to analogous effects. The answer to this question is unclear since for magnetic fields with a coherence length higher than $R_{\rm core} \gtrsim \left(g\, B' \right)^{-1/2}$ (and uncorrelated on such length scales), the Higgs origin of the gauge boson mass makes the localization of the magnetic field inside vortices energetically favourable~\cite{Adelberger:2003qx}. To our knowledge, no study on the evolution of such scenario has been performed. Furthermore, the existence of a non-negligible coherent magnetic field on such long scales could hardly be sustained due to the strong pressure sourced by the mass term in the Proca action~\cite{Goldhaber:1971mr,Adelberger:2003qx}.

\section{Results}
\label{sec:results}

\begin{figure}
    \centering
    \includegraphics[width = 0.9\linewidth]{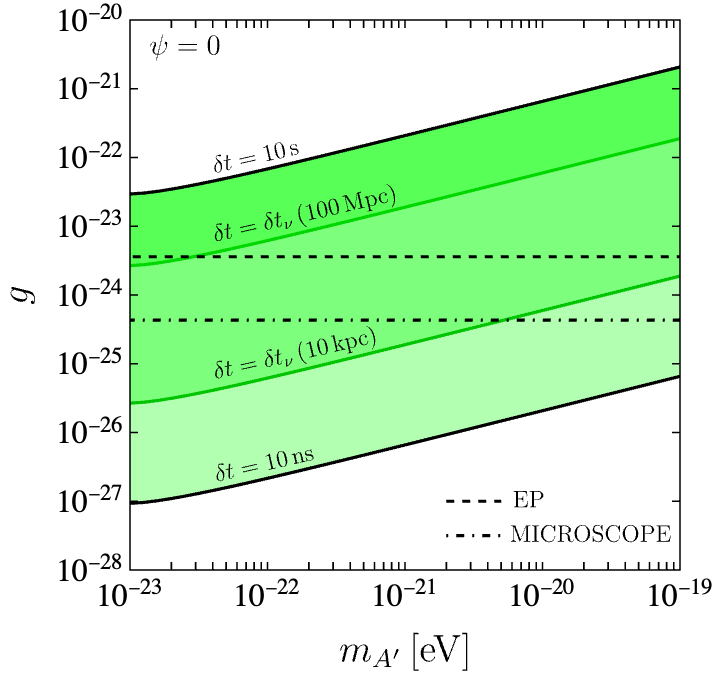}
    \caption{The portion of parameter space that can be probed by future measurements of a neutrino flux of average energy 15\,MeV from a distant supernova coming from the direction at the opposite from the Solar system in the Milky Way, $\psi=0$, as a function of the gauge boson mass $m_{A'}$ in eV (horizontal axis) and the coupling $g$ (vertical axis). Different shades of green denote different delays in the arrival timings compared to the photon signals. The shaded areas within the thick lines show the region of the parameter space that is allowed given the time delay $\delta t = 10$\,s which is taken as a reference (upper bound) and the reach which is expected to be probed by the DUNE collaboration, assuming a sensitivity $\delta t = \mathcal{O}(10{\rm\,ns})$ in the time delay tagging (lower bound). The two intermediate lines mark the regions where the delay due to the finite mass of the neutrino masks the effect of the gauge boson for a supernova placed at $d=100\,$Mpc or $d=10\,$kpc. Also shown are the current upper bounds placed by the E\"ot-wash experiment (labeled ``EP'') and the MICROSCOPE experiment, see the text for details.}
    \label{fig:mgdependence}
\end{figure}

We collect the relevant information on the parameter space allowed within the theory in Fig.~\ref{fig:mgdependence}, as a function of the gauge boson mass $m_{A'}$ in eV and the coupling $g$. The bounds are shown for a neutrino burst of average energy 15\,MeV reaching Earth from a region at the opposite end of the Galactic center, set by an angle $\psi = 0$. We remind that the angle of incidence must satisfy the relation in Eq.~\eqref{eq:psiangleconstr} for the neutrino burst to cross the galactic soliton core and the bending effect to take place. For other values of $\psi$ obeying the constraint in Eq.~\eqref{eq:psiangleconstr}, the plot remains qualitatively unchanged. Given the results presented in Fig.~\ref{fig:mgdependence}, some considerations are in place.

We account for the bounds from supernova neutrinos, working under the assumption that the density of the soliton core expressed in Eq.~\eqref{eq:rhosoliton} tracks the galactic distribution within the radius $R_{\rm core}$. To frame the parameter space for the expected neutrino time delay in Eq.~\eqref{eq:timedelay}, we have adopted the upper bound on the neutrino burst released during the SN1987A event, $\delta t \lesssim 10$\,s~\cite{Arnett:1987iz, Kolb:1987dda, Raffelt:1996wa}. This leads to the upper bound on the gauge boson coupling in Eq.~\eqref{fig:mgdependence}. However, present-day data acquisition allows to reach a much lower sensitivity in the time delay $\delta t \sim \mathcal{O}(10{\rm\,ns})$, such as what is employed for the prototype of the DUNE facility~\cite{Belver:2021drc, DUNE:2022ctp} and leading to the lowest bound on the coupling $g$ that can be attained in the laboratory for the model.

The effects on the time delay from the bending of the neutrino trajectories becomes comparable to an analogous time delay due to the finite neutrino mass $\delta t_\nu$ obtained in Eq.~\eqref{eq:distancenu}. For example, this delay amounts to $\delta t_\nu \approx 0.1$\,s when considering a supernova at a distance $d = 100\,$Mpc, or $\delta t_\nu \approx 10\,\mu$s when $d = 10\,$kpc. These time lags lie above the sensitivity of DUNE discussed above and, in the absence of multiple events, lead to a lower bound on the gauge boson coupling in the case of a detection. Nevertheless, as mentioned below Eq.~\eqref{eq:distancenu}, the two effects in Eqs.~\eqref{eq:timedelay} and~\eqref{eq:distancenu} can be differentiated if the neutrino masses are measured with a sufficient accuracy or if multiple supernovas are detected.

An absolute lower bound on the gauge coupling is suggested from considerations over the weak gravity conjecture~\cite{Arkani-Hamed:2006emk}, which would require $g \gtrsim m_\nu/m_{\rm Pl} \sim 10^{-30}$, for a neutrino mass $m_\nu \sim 10^{-2}\,$eV and where $m_{\rm Pl} = 1/\sqrt{G}$. This value is several orders of magnitudes below the smallest values of the coupling that can be probed in the model proposed, hence we can safely ignore the effects of such a conjecture in its weakest form.

In Fig.~\ref{fig:mgdependence}, we have not shown the loose upper bound on the gauge coupling from the screening effect of the cosmic neutrino background previously discussed around Eq.~\eqref{eq:gblscreening}, which  amounts to $g \lesssim 10^{-16}$. Similarly, we do not show the lines resulting from constant values of the deflection angle $\Theta$ in Eq.~\eqref{eq:Theta}, since such an angle is not measurable in the parameter space of interest. For reference, we report the bound from torsion balance experiments such as the E\"ot-Wash experiment~\cite{Wagner:2012ui}, placing an upper bound on the coupling $g \lesssim 3.6 \times 10^{-24}$ at 95\% confidence level (CL) for ultralight masses (dashed line). Recently, the MICROSCOPE experiment~\cite{Touboul:2017grn, MICROSCOPE:2022doy} placed an even stronger bound $g \lesssim 4.3 \times 10^{-25}$ at 95\% CL in the same mass window (dot-dashed line).\footnote{See also Refs.~\cite{Berge:2017ovy, Fayet:2017pdp, Fayet:2018cjy} for the derivation of the bounds.} These bounds, however, only apply to the case of a $B\textrm{--}L$ gauge symmetry and are not applicable, e.g., to the $L_{\mu}\textrm{--}L_{\tau}$ symmetric model discussed in Sec.~\ref{sec:methods}.

Some considerations are in place. Within an explicit UV completion, the related symmetry breaking scale $v'$ in the parameter space of interest is of the order of $v'\sim\mathcal{O}(\textrm{eV}- \textrm{MeV})$. This implies the existence of a light sterile neutrino $N$ of similar mass. In such a scenario, to avoid the stringent constraints on the number of active neutrino species at recombination $N_{\rm eff}$, the sterile neutrino mass should be $m_{N}\sim\,$eV with the mixing angles smaller than $\sim 10^{-1}$~\cite{Vagnozzi:2019ezj}, see also the recent review on the constraints on light sterile neutrinos in Ref.~\cite{Bolton:2019pcu}. An oscillating ultralight bosonic field, either in the form of a scalar or a vector field, could potentially reveal itself as a time modulation over the effective neutrino mass. This can be explored efficiently with the upcoming neutrino laboratories such as JUNO~\cite{JUNO:2015zny} and DUNE~\cite{DUNE:2015lol}. This effect has been explored for an ultralight scalar~\cite{Krnjaic:2017zlz, Brdar:2017kbt, Dev:2020kgz} as well as for light vector gauge bosons~\cite{Brdar:2017kbt, Alonso-Alvarez:2023tii}. A different effect of the light gauge boson as the DM consists in the modulation of the sterile neutrino mass, thus evading cosmological bounds~\cite{Davoudiasl:2023uiq}.

\section{Discussion}
\label{sec:discussion}

Previous work has considered the time delay of SN neutrinos resulting from their interactions with other sources. While the bending in the trajectory is specific to this model, new interactions with dark matter (DM) might further delay the propagation of neutrinos due to scattering, leading to a diffused propagation characterized by the root-mean-square scattering angle $\theta_{\rm rms}$, which corresponds to a time delay~\cite{Murase:2019xqi, Eskenasy:2022aup}:
\begin{equation}
    \label{eq:deltatBSM}
    \delta t_{\rm BSM} \simeq \frac{1}{2}\,\frac{\theta_{\rm rms}^2}{4}\,d\,.
\end{equation}
For example, vector-mediated interactions between neutrinos and fermionic DM could lead to a significant time delay of SN neutrinos, accumulating over the propagation length of the neutrino beam~\cite{Carpio:2022sml}. Beyond new interactions, a time delay between neutrinos and photons could arise from violations of the weak equivalence principle or Lorentz invariance~\cite{Wang:2016lne, Ellis:2018ogq, Boran:2018ypz, Laha:2018hsh, Wei:2018ajw}. However, the effect we discussed occurs only within the solitonic core, as it relies on the coherent motion of the ultralight vector field, effectively replacing Eq.~\eqref{eq:deltatBSM} with Eq.~\eqref{eq:timedelay}. Moreover, in the model we considered, the neutrino beam must traverse the solitonic core according to Eq.~\eqref{eq:psiangleconstr}, meaning that any future detection of a significant time delay from different directions would be a clear signal of new physics unrelated to the model considered.

Neutrino self-interactions mediated by new bosons would result in an additional delay in their free-streaming when active at temperatures above neutrino decoupling~\cite{Cyr-Racine:2013jua, Archidiacono:2013dua, Lancaster:2017ksf, Oldengott:2017fhy, Kreisch:2019yzn}, contributing to an increased $N_{\rm eff}$~\cite{Blinov:2019gcj, Brinckmann:2020bcn}. These interactions can also impact multimessenger searches by influencing the interactions of neutrinos with the cosmic neutrino background and dark matter (DM)~\cite{Murase:2019xqi}. Previous work has explored a vector mediator between neutrinos and fermionic DM $\chi$~\cite{Carpio:2022sml}, which involves two couplings: $g_{\nu A'} \equiv g$ and $g_{A'\chi}$. In our model, we do not include the additional coupling $g_{A'\chi}$ since DM is not fermionic. However, we expect a similar effect due to neutrino quartic interactions or a coupling between the gauge boson and electrons. For a quartic self-interaction mediated by a gauge boson, the value of $g$ that can be probed based on the results of Fig.~\ref{fig:mgdependence} is well below current experimental constraints~\cite{AristizabalSierra:2020edu, Boehm:2020ltd}. Moreover, this scenario could be tested at neutrino facilities. For instance, if DM consists of light bosons, its coupling with solar neutrinos could alter the patterns of neutrino flavor oscillations. This effect has not been observed in the Sudbury Neutrino Observatory (SNO)\cite{SNO:2011hxd} or Super-Kamiokande\cite{Super-Kamiokande:2001ljr}, placing a constraint on the scalar-neutrino Yukawa coupling~\cite{Berlin:2016woy}.

In addition to the abundance of cold gauge bosons, other messengers can be searched jointly with the bending of the neutrino trajectories to pin down the scenario that has led to the DM formation. If, as outlined above, the DM population forms from parametric resonance, there would be indication from the specific model used such as the scale of inflation~\cite{Bastero-Gil:2018uel} or the presence of a new pseudoscalar field~\cite{Agrawal:2018vin}. In case of an exotic production of the gauge bosons from a phase transition or cosmic string network dynamics~\cite{Vachaspati:1984gt, Vilenkin:1986ku}, the concurrent release of GWs would contribute to a background in the present Universe with a characteristic spectrum that can be searched along with the effect proposed here. In the context of a $L_{\mu}\textrm{--}L_{\tau}$ symmetry-breaking theory, the ultralight gauge boson dark matter in the parameter space of Fig.~\ref{fig:mgdependence} can mimic the GW signal detected at the nHz range~\cite{Chowdhury:2023xvy}.

We have considered SN neutrinos as a test case, given that their spectrum is well understood, and the quoted average energy of 15 MeV serves as a benchmark in the literature. Other highly-coherent neutrino sources could also be viable, especially when a multimessenger test can be performed. For instance, binary neutron star mergers could produce joint signals detectable by next-generation gravitational wave detectors, along with high-energy neutrino searches at future neutrino observatories~\cite{Mukhopadhyay:2023niv, Mukhopadhyay:2024lwq}. Another instance of joint observation is the detection of a single neutrino from the flaring blazar TXS 0506+056~\cite{IceCube:2018dnn}, which suggests that blazars may be sources of high-energy neutrinos. Both binary mergers and blazars are typically located at extragalactic distances, far greater than those of galactic supernovae, leading to increased uncertainties in energy and timing resolution. However, the scenario we propose avoids these drawbacks, as neutrino bending occurs only within the galactic core and does not otherwise affect neutrino propagation.

We further comment on the choice of the soliton solution in Eq.~\eqref{eq:rhosoliton}. While this solution has been confirmed in independent searches~\cite{Veltmaat:2018dfz}, cosmological simulations of ultralight (bosonic) DM admixed with cold DM or baryons generally predict that the soliton would shrink while becoming more massive and denser~\cite{Veltmaat:2019hou, Lague:2023wes}. A smaller soliton is expected as the mass of the ultralight vector field increases, with $R_{\rm core} \lesssim 100$\,pc for $m_{A'}\gtrsim 10^{-21}$\,eV. Nevertheless, for a $B\textrm{--}L$ gauge symmetry this region has already been probed in laboratory experiments, see Fig.~\ref{fig:mgdependence}. On the other hand, a denser distribution of the solitonic region would lead to an enhanced bending.

A further improvement in the modeling proposed that leads to a more realistic approach would account for a series of modifications in the computation sketched concerning the distribution of the DM in the Galaxy. Given the spatial coherence length at the onset of field oscillations $\lambda_{\rm in}\sim m_{A'}^{-1}$, the present-day correlation length $\lambda_0 = \lambda_{\rm in}(a_0/a_{\rm in})$ gives the spatial variation over which the space-averaged gauge boson field oscillates with the frequency $m_{A'}$. Along with these coherent patches of correlation length $\lambda_0$, other structures such as ``mini-voids'' and filaments are expected to form~\cite{Eggemeier:2022hqa}. These dense patches are also expected to experience disruption from massive regions such as galaxies and clusters, possibly depleting their abundance~\cite{Kavanagh:2020gcy}. This is an impactful line of search that is left for future exploration.

\section{Conclusions}
\label{sec:conclusions}

We have studied the consequences of a cosmic abundance of massive gauge bosons with a transverse component on the propagation of galactic neutrinos. An ultralight gauge boson as the DM organizes itself as a coherent solitonic core over a length scale $R_{\rm core}$, with its transverse component adjusted to an effective electric field with a random orientation. A neutrino propagating in the galactic core for a distance $L$ would experience a total angular deflection and a time delay with respect to the straight propagation because of the action of the coupled gauge field. We have studied these effects in a general scenario that does not account for the specific DM production mechanism, bearing in mind some realistic models in which the transverse component is realized~\cite{Dror:2018pdh, Bastero-Gil:2018uel, Agrawal:2018vin, Kitajima:2022lre, Amaral:2024tjg}.

We find that the model proposed might result in a time delay of a neutrino beam from a nearby supernova in the range $\delta t = \left(10^{-8}\textrm{--}10^1\right)$\,s for the gauge boson masses in the ``fuzzy" DM range, $m_{A'} = \left(10^{-23}\textrm{--}10^{-19}\right)$\,eV and the coupling with the neutrino $g= 10^{-27}\textrm{--}10^{-20}$. In this range of parameters, a little seesaw mechanism for the generation of neutrino mass is possible, see e.g.\ Refs.~\cite{Kohri:2008cf, Cabrera:2023rcy}.

%\vspace{0.5cm}

\section*{Acknowledgments}
We thank Andrea Caputo for reviewing a preliminary version of the draft and for insightful comments. We additionally thank Sergio Bertolucci for the discussion over the data acquisition timing at DUNE. L.V.\ and M.Z.\ acknowledge support by the National Natural Science Foundation of China (NSFC) through the grant No.\ 12350610240 ``Astrophysical Axion Laboratories'', as well as hospitality by the Istituto Nazionale di Fisica Nucleare (INFN) Frascati National Laboratories. L.V.\ additionaly thanks the INFN section of Trento (Italy), the INFN section of Ferrara (Italy), and the University of Texas at Austin (USA) for the support. This publication is based upon work from the COST Actions ``COSMIC WISPers'' (CA21106) and ``Addressing observational tensions in cosmology with systematics and fundamental physics (CosmoVerse)'' (CA21136), both supported by COST (European Cooperation in Science and Technology). T.T.Y.\ also thanks the NSFC through the grant No.\ 12175134, JSPS Grant-in-Aid for Scientific Research Grants No.\ 24H02244, and World Premier International Research Center Initiative (WPI Initiative), MEXT, Japan.

%%%%%%%%%%%%%%%%%%%%%%%%%%%%%%%%%%%%%%%%%%%%%%
%\setlength{\bibsep}{4pt}
\bibliographystyle{elsarticle-num}
\bibliography{citations}
\end{document}